\documentclass[aps,pra,twocolumn,groupedaddress]{revtex4} 
\usepackage{graphicx}
\usepackage{epstopdf}
\usepackage{bm}

\begin{document}

\title{Repumping and spectroscopy of laser-cooled Sr atoms using the $(5s5p) ^3$P$_2$ - $(5s4d) ^3$D$_2$ transition}
\author{P. G. Mickelson, Y. N. Martinez de Escobar, P. Anzel, B. J. DeSalvo, S. B. Nagel, A. J. Traverso, M. Yan, T. C. Killian}
\affiliation{Rice University, Department of Physics and
Astronomy, Houston, Texas, 77251}

\date{\today}

\begin{abstract}
We describe repumping and spectroscopy of laser-cooled strontium (Sr) atoms using the $(5s5p) ^3$P$_2$ - $(5s4d) ^3$D$_2$ transition.  Atom number in a magneto-optical trap is enhanced by driving this transition because Sr atoms that have decayed into the $(5s5p) ^3$P$_2$ dark state are repumped back into the $(5s^2) ^1$S$_0$ ground state.  Spectroscopy of $^{84}$Sr, $^{86}$Sr, $^{87}$Sr, and $^{88}$Sr improves the value of the $(5s5p) ^3$P$_2$ - $(5s4d) ^3$D$_2$ transition frequency for $^{88}$Sr and determines the isotope shifts for the transition. 
\end{abstract}


\maketitle

Cold atom experiments require cycling transitions for efficient laser cooling and trapping.  Depending on the level structure, atoms may be shelved into dark states during laser cooling, which removes them from the cooling cycle and can cause them to be lost from the trap.  By applying laser light of the appropriate frequency to shelved atoms, it is possible to return these atoms to the cycling transition \cite{Neuhauser:1978,Phillips:1985}.  This repumping process can increase atom number and density, which improves signal-to-noise ratios for most measurements, enables study of collisional processes, and is crucial for achieving quantum degeneracy \cite{Ketterle:1993}.

In experiments with alkali-metal atoms, the dark states are ground state hyperfine levels, and repumping lasers can be generated with acousto-optic or electro-optic modulators from the laser used for cooling.  In alkaline-earth-metal atoms such as strontium (Sr), however, atom population is trapped in highly excited metastable levels and independent lasers are necessary.  Despite requiring additional lasers, alkaline-earth-metal atoms are interesting to study because they offer the possibility of an all-optical path to quantum degeneracy \cite{Katori:1999qb, Ido:2000kb}, possess narrow optical transitions that can be used for optical frequency standards \cite{Ye:2008}, and provide fine control of atomic interactions via optical Feshbach resonances \cite{Ciurylo:2004qo}.

\begin{figure}
\centering
\includegraphics[clip=true,keepaspectratio=true,width=3in,height=3in]{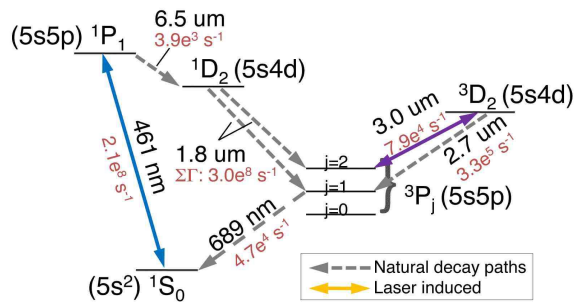}\\
\caption[Level diagram depicting repumping scheme]{Wavelengths and decay rates for selected Sr transitions.  The OPO laser enables the repumping scheme outlined in the text by pumping atoms that have leaked from the $^1$P$_1$ to the $^3$P$_2$ state up to the $^3$D$_2$ state, thus allowing decay to the $^3$P$_1$ state and subsequent return to the $^1$S$_0$ ground state.  The main cycling transition operates on the $^1$S$_0$ to $^1$P$_1$ transition, and time-of-flight absorption imaging of ground state atoms is performed using 461 nm light.
\label{leveldiagram}
}
\end{figure}

For Sr, the principal cycling transition for laser cooling operates between the $(5s^2) ^1$S$_0$ and the $(5s5p) ^1$P$_1$ states (Fig. \ref{leveldiagram}).  Decay via the $(5s5p) ^1$P$_1$ - $(5s4d) ^1$D$_2$ transition \cite{Loftus:2002} allows atoms to escape the cycling transition, and further decay from the $(5s4d) ^1$D$_2$ state results in atoms in the $(5s5p) ^3$P$_1$ and $(5s5p) ^3$P$_2$ states (henceforth $^3$P$_j$).  $^3$P$_1$ atoms return to the ground state and are recaptured in the MOT, but $^3$P$_2$ atoms are shelved because of the 17 min lifetime of the $^3$P$_2$ state \cite{Derevianko:2001}.

Here, we describe a repumping scheme for Sr using the $^3$P$_2$ - $(5s4d) ^3$D$_2$ transition at 3012 nm which has a historically difficult-to-reach wavelength in the mid-infrared (MIR).  Lasers of this frequency based on optical parametric oscillators have recently become available due to advances in nonlinear optics and fiber lasers.  Among the advantages this transition offers is the simplicity it brings in comparison to repumping schemes like the one described in \cite{Nagel:2003ff} or \cite{Xu:2003uq}.  Similar transitions have been used to create a calcium MOT \cite{Grunert:2001} operating on the 1978 nm $^3$P$_2$ - $^3$D$_3$ cycling transition and in another Sr experiment \cite{Poli:2005qf} that uses the $(5s5d) ^3$D$_2$ for the upper level, with a transition wavelength of 496 nm, to repump atoms out of the $^3$P$_2$ state.

We also determine an improved value of the transition frequency and perform spectroscopy of the $^3$P$_2$ - $(5s4d) ^3$D$_2$ transition for $^{84}$Sr, $^{86}$Sr, $^{87}$Sr, and $^{88}$Sr.  Using these spectra, we assign isotope shifts for the $^{84}$Sr, $^{86}$Sr, and $^{87}$Sr transition relative to the $^{88}$Sr transition.

Our experiment begins similarly to previously published work \cite{Nagel:2003ff,Nagel:2005pk,Mickelson:2005qw}.  As many as 50 x 10$^6$  $^{88}$Sr atoms are trapped in a magneto-optical trap (MOT) operating on the 461 nm cycling transition between the $^1$S$_0$ and the $^1$P$_1$ states.  The MOT beams, red-detuned by 60 MHz from resonance and with intensity-per-beam $I $= 2.3 mW/cm$^{2}$, yield atom samples with a temperature of about 2 mK, a density on the order of 10$^{10}$ cm$^{-3}$, and a $1/e$ radius of about 1 mm.  For spectroscopy, we also trap other Sr isotopes \cite{Kurosu:1990}, $^{84}$Sr ($<$ 1 x 10$^6$ atoms), $^{86}$Sr (10 x 10$^6$ atoms), and $^{87}$Sr (5 x 10$^6$ atoms).  Light at 461 nm is produced by frequency doubling via KNbO$_3$ in a linear enhancement cavity \cite{Bode:1997fd}.  Time-of-flight absorption imaging is also performed using the  $^1$S$_0$ - $^1$P$_1$ transition.

We produce 3 $\mu$m light for repumping and spectroscopy using a laser based on optical parametric oscillation (OPO) which is seeded by a fiber laser at 1.06 $\mu$m \cite{Henderson:2006}.  Our experiments only require a minimal amount of power, typically about 4 mW incident on the atoms, and the beam has a $1/e^2$ radius of about 3 mm.  We frequency-stabilize the laser to 0.002 cm$^{-1}$ precision using a calibrated wavemeter.

\begin{figure}
\centering
\includegraphics[clip=true,keepaspectratio=true,width=3in,height=3in]{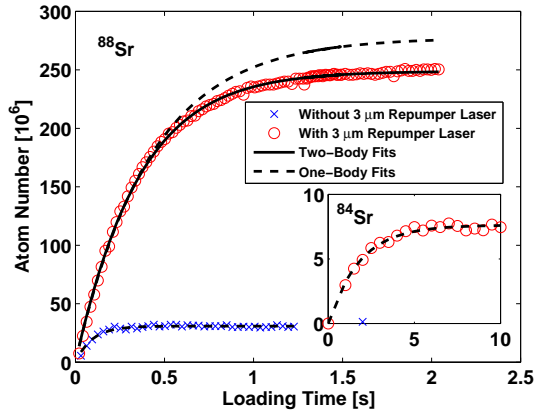}\\
\caption[MOT loading curves with and without repumper]{Here we show the number of $^{88}$Sr and $^{84}$Sr (inset) atoms trapped as a function of time with and without application of mid-infrared (MIR) laser light at 3 $\mu$m. Without the MIR light, atoms excited by the cooling laser that decay to the metastable $^3$P$_2$ state are lost from the trap, which limits the maximum number. The MIR laser pumps the metastable atoms to a state that decays back to the ground state so that they are not lost.  Using the model described in the text, we determine that two-body collisions are limiting the maximum number of $^{88}$Sr atoms when the repumping laser is on; a one-body fit to early-time data (dashed line) overestimates the final number when the repumper is on, whereas it is a good fit when the repumper is off.  Only 300,000 $^{84}$Sr atoms are observed without the repumper because the natural abundance of $^{84}$Sr (0.56$\%$) is very low.  The enhancement of $^{84}$Sr due to the repumper is larger than that of $^{88}$Sr primarily because of improved vacuum conditions during the $^{84}$Sr experiment. 
\label{motloadingcurves}
}
\end{figure}

As described earlier, the cycling transition used for the MOT is not closed because of leakage from the $^1$P$_1$ state, leading to shelving of atoms in the $^3$P$_2$ state.  Figure \ref{motloadingcurves} shows the number of atoms as a function of the MOT loading time with and without the repumping laser applied.  Absent the repumping laser, the atom number is significantly lower than when the repumping laser enables a return path to the ground state.

We examine the enhancement the repumping laser brings to the steady-state number of atoms using the time-dependent number equation for MOT loading:
\begin{equation}
\dot{N} = L_N - \Gamma N - \beta^{\prime} N^2.
\label{eqnrates}
\end{equation}
Here, $N$ is the number of atoms, $L_N$ is the loading rate of atoms into the MOT, $\Gamma$ is the one-body loss rate, and $\beta^{\prime} = \beta/(2\sqrt{2}V)$, where $\beta$ is the two-body loss constant and $V = \int d^3r e^{-\frac{r^2}{\sigma^2}}$ is the effective volume for two-body processes ($\sigma$ is the 1/$\sqrt{e}$ radius and $r$ is position).  The solution to this differential equation is
\begin{equation}
N(t)=\frac{N_{ss}(1-e^{-\gamma t})}{(1+\chi e^{-\gamma t})},
\end{equation}
with $\gamma=\Gamma+2\beta^{\prime}N_{ss}$, $N_{ss}$ the steady state number of atoms, and $\chi$ the measure of the relative contributions of the one- and two-body loss coefficients:
\begin{equation}\label{nssequation}
N_{ss}=\frac{-\Gamma+\sqrt{\Gamma^2+4\beta^{\prime}L}}{2\beta^{\prime}}
\end{equation}
and
\begin{equation}\label{chiequation}
\chi=\frac{\beta^{\prime} N_{ss}}{\beta^{\prime} N_{ss}+\Gamma}.
\end{equation}
Using this model, we determine the fits shown in Fig. \ref{motloadingcurves}.  Without the repumping laser, a one-body fit (dashed line), with $\beta^{\prime} = 0$ and $\Gamma$ = 10.7$\pm$0.5 s$^{-1}$ is consistent with optical pumping of atoms to the $^3$P$_2$ state by the MOT laser \cite{Dinneen:1999fb}.  A two-body fit, with $\beta$ = 6$\pm$2 $\times$10$^{-11}$ cm$^3$/s and $\Gamma$ = 2.4$\pm$0.1 s$^{-1}$, fits the data when the repumping laser is on.  A one-body fit to the first 0.5 s of this data (dashed line), overestimates the steady state number of atoms.  This value of $\beta$ is only approximate, as care was not taken to accurately measure the sample volume, V, but it indicates that two-body processes are limiting the number of atoms loaded into the MOT.  $\beta$ is slightly lower than the value found in \cite{Dinneen:1999fb} which is reasonable given the larger detuning of our MOT laser frequency and the lower intensity of our MOT beams.

\begin{figure}
\centering
\includegraphics[clip=true,keepaspectratio=true,width=3in,height=3in]{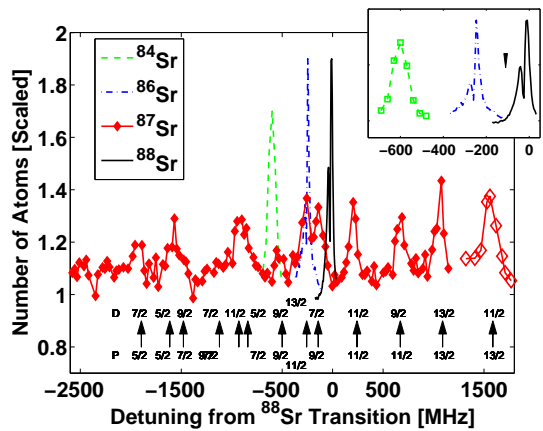}\\
\caption[Repumping spectrum]{Spectroscopy of the $^3$P$_2$ - $^3$D$_2$ transition.  Shifts are measured relative to the zero of the $^{88}$Sr spectrum.  For $^{86}$Sr, $^{87}$Sr, and $^{88}$Sr, the number is normalized to the number observed without the repumping laser.  For $^{84}$Sr, the scaling is arbitrary because large repumping efficiency is necessary to observe a spectrum.  Structure in the $^{87}$Sr spectrum is due to the hyperfine interaction: the fermionic isotope of Sr has nuclear spin $I$ equal to 9/2.  Level assignments (arrows) can be made for all of the observed peaks, and the isotope shift of $^{87}$Sr is determined by the shift of the centroid of the energy level manifold from the $^{88}$Sr zero.  Inset: the arrow in the inset shows the position of the centroid for the $^{87}$Sr hyperfine levels.  The structure observed in the $^{86}$Sr and $^{88}$Sr lines is due to Zeeman splitting caused by the 50 G/cm magnetic field gradient of the MOT magnetic field.  The gradient also contributes some broadening to the lines.  Structure is not resolved for $^{84}$Sr and $^{87}$Sr because the spectra are observed only at high laser power, which washes out the structure.
\label{spectrumSr}
}
\end{figure}

Using the repumping of atoms, we performed spectroscopy of the $^3$P$_2$ - $^3$D$_2$ transition for all the stable isotopes of Sr.  For this study, we observe the repumping enhancement in the steady-state number of MOT atoms, although trapping of $^3$P$_2$ atoms in the magnetic trap formed by the quadrupole magnets of the MOT \cite{Nagel:2003ff} can affect the results.  Scanning the laser across the resonance frequency of the repumping transition changes the number of atoms imaged (Fig. \ref{spectrumSr}).  The structure of the even isotopes, $^{84}$Sr, $^{86}$Sr, and $^{88}$Sr, is simpler than that of the odd isotope, $^{87}$Sr, because the even isotopes have nuclear spin equal to zero.

At low repumping laser intensity, the spectra of $^{86}$Sr and $^{88}$Sr (see inset of Fig. \ref{spectrumSr}) reveal structure arising from Zeeman splitting due to the 50 G/cm magnetic field gradient of the MOT magnetic coils.  The detailed dynamics of the repumping process are beyond the scope of this paper. We suspect that at the low repumper intensities used for these isotopes, the repumping is slow enough that atoms escape the region of the MOT unless they are in the $m_j=2$ and $m_j=1$ sublevels and are magnetically trapped \cite{Nagel:2003ff}, and $m_j=2$ is more populated because it is trapped more strongly.  The double peaks we observe are likely due to transitions from the m=2 state of $^3$P$_2$ to the m=2 and m=1 states in the $^3$D$_2$ manifold.  The observed splitting matches what one would expect from the known magnetic moments of the upper and lower levels, the magnetic field gradient, and the temperature of atoms in the MOT \cite{Nagel:2003ff}.  This simple model allows us to determine the position of the unperturbed resonances (Fig. \ref{spectrumSr} inset).  For $^{84}$Sr and $^{87}$Sr,  all the repumping laser power is necessary to achieve signal because of the low natural abundance of $^{84}$Sr (0.56 $\%$) and the poor repumping efficiency of $^{87}$Sr, and no structure is observed.  For these isotopes, the unperturbed resonances are taken as the center of the line.

The $^{87}$Sr spectrum shows hyperfine structure because it has a nuclear spin of $I$=9/2, but since the spectra are taken at high repumping laser intensity, no magnetic sublevels are observed.  We calculate the positions of the hyperfine states using the Casimir formula:
\begin{equation}
\Delta E_F = A \frac{K}{2} + B \left[\frac{3/4 K(K+1)-I(I+1)J(J+1)}{I(2I-1)2J(2J-1)}\right],
\label{casimirformula}
\end{equation}
with $K=F(F+1)-J(J+1)-I(I+1)$ and the values of the magnetic dipole and electric quadrupole factors ($A$ and $B$, respectively) taken from \cite{Bushaw:1993} for the $^3$D$_2$ level and from \cite{Heider:1977} for the $^3$P$_2$ level.  For this transition $J = 2$ and $I = 9/2$, and the total angular momentum, $F$, varies from $5/2$ to $13/2$ for both the upper and lower states of the transition.  We overlay the calculated positions on the observed spectrum to assign the experimental peaks to the calculated positions.

\begin{figure}
\centering
\includegraphics[clip=true,keepaspectratio=true,width=3in,height=3in]{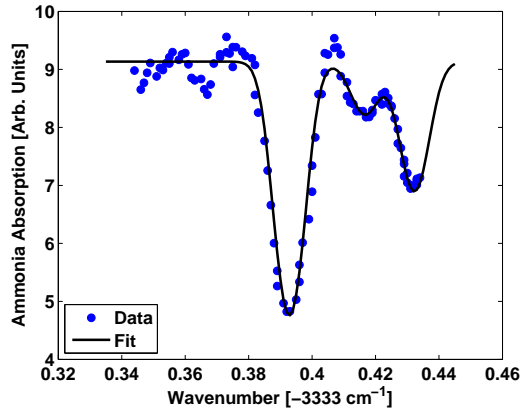}\\
\caption[Wavemeter calibration]{Absorption spectroscopy of ammonia for wavemeter calibration.  We fit the peaks using a multiple Gaussian line shape and compare the center frequency of the strongest line to data from \cite{Guelachvili:1989} to determine the systematic error of our wavemeter readings.  The uncertainty of our fit to these frequencies is about 0.0015 cm$^-1$.
\label{wavemetercal}
}
\end{figure}

To calibrate the wavemeter absolutely, we perform absorption spectroscopy of ammonia in a gas cell at room temperature and $\sim$1 torr (Fig. \ref{wavemetercal}).  Expected pressure shifts on the order of one MHz \cite{Bernath:1995} are negligible.  An accurate wavelength value for the strong peak (Table \ref{HITRANtable}) can be found in \cite{Guelachvili:1989}, which allows determination of the systematic offset of the wavemeter measurement (-0.004 cm$^-1$).  We correct for the systematic error in our wavemeter when stating our measurements of the Sr transition.  We find the resonance wave number of the $^3$P$_2$ - $^3$D$_2$ transition in $^{88}$Sr to be 3320.226$\pm$0.0025 cm$^{-1}$, which is a small shift and improvement over the previously available value of 3320.232 cm$^{-1} $\cite{Sansonetti:2005}.  Our uncertainty arises from statistical uncertainty in fitting the lines in Fig. \ref{wavemetercal} and from drifts in the wavemeter calibration.  

\begin{table}[htdp]
\caption{Wavemeter calibration with ammonia absorption line.}
\begin{center}
\begin{tabular}{|c|c|}
\hline
Observed Level [cm$^{-1}$] & Ref. \cite{Guelachvili:1989} Level [cm$^{-1}$] \\
\hline
3333.3928(15) & 3333.3975(10) \\
\hline
\end{tabular}
\end{center}
\label{HITRANtable}
\end{table}

\begin{table}[htdp]
\caption{Isotope shifts and uncertainties of the $^3$P$_2$ - $^3$D$_2$ transition at $\lambda$=3012 nm in Sr.}
\begin{center}
\begin{tabular}{|c|c|}
\hline
Isotope Pair & Isotope Shift [MHz] at $\lambda$ \\
\hline
87-88 & -110(30) \\
86-88 & -270(40) \\
84-88 & -600(50) \\
\hline
\end{tabular}
\end{center}
\label{IStable}
\end{table}

\begin{figure}
\centering
\includegraphics[clip=true,keepaspectratio=true,width=3in,height=3in]{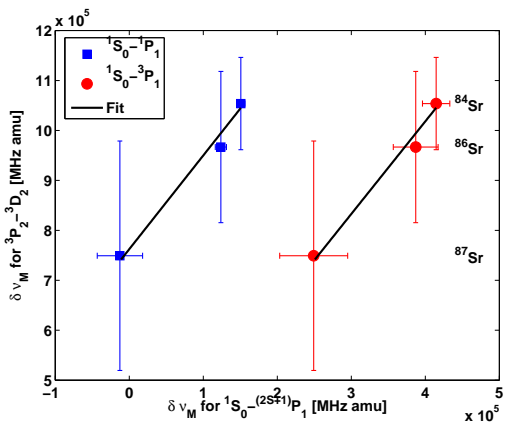}\\
\caption[King Plot for Isotope Shifts]{A King plot of the modified isotope shifts, $\delta \nu_M$, of the $^3$P$_2$ - $^3$D$_2$ transition versus the modified isotope shifts of the 461 nm $^1$S$_0$ - $^1$P$_1$ \cite{Bushaw:2000} and 689 nm $^1$S$_0$ - $^3$P$_1$ \cite{Bushaw:1997} transitions of Sr.
\label{kingplot}
}
\end{figure}

Table \ref{IStable} lists the isotope shifts relative to $^{88}$Sr.  The uncertainties reflect uncertainty in fitting and modeling the lines.  Figure \ref{kingplot} compares our values for the isotope shifts to previous isotope shift measurements on the $^1$S$_0$ - $^1$P$_1$ \cite{Bushaw:2000} and $^1$S$_0$ - $^3$P$_1$ \cite{Bushaw:1997} Sr lines with a King plot \cite{King:1963,Dammalapati:2008} of the modified isotope shift ($\delta \nu_M$), 
\begin{equation}
\delta \nu_M = (\delta \nu_{IS} - \delta \nu_{NMS}) \frac{A_1A_2}{A_1-A_2},
\label{modisotopeshift}
\end{equation}
where $A_1$ and $A_2$ are the mass numbers in atomic mass units (amu) of the isotopes, $\delta \nu_{IS}$ is the observed isotope shift, and $\delta \nu_{NMS} = (\nu m_e/m_p) \times (A_1-A_2)/A_1A_2$ is the normal mass shift caused by the reduced mass of the atom ($\nu$ is the frequency of the transition; $m_e$ and $m_p$ are electron and proton masses).  Within the error, this King plot shows the expected linear relations between the isotope shifts for the different transitions.

In conclusion, we have shown repumping of all stable isotopes of Sr using the $^3$P$_2$ - $^3$D$_2$ transition.  Additionally, we have measured the isotope shift of the $^3$P$_2$ - $^3$D$_2$ transition for $^{84}$Sr, $^{86}$Sr, and $^{87}$Sr and provided an improved value for the $^3$P$_2$ - $^3$D$_2$ transition wavelength of $^{88}$Sr.

\end{document}